\documentclass[twocolumn,amssymb,amsmath,aps,showpacs,nofootinbib,nobibnotes,pra]{revtex4-1}

\usepackage{graphicx}
\usepackage{xcolor}
\usepackage[colorlinks=true,linkcolor=blue,urlcolor=blue,citecolor=blue,bookmarks=true,bookmarksopenlevel=2]{hyperref}
\usepackage{amsmath}

\usepackage[utf8]{inputenc}
\usepackage[T1]{fontenc}
\usepackage{microtype}

\newcommand{\alert}[1]%
{%
\textcolor{red}{#1}
}%

\newcommand{\ket}[1]{\left\vert#1\right\rangle}

\newcommand{\sandwich}[3]{\left\langle #1 \vphantom{#2 #3} \right\vert #2 \left\vert \vphantom{#1 #2} #3 \right\rangle}

\begin{document}

\title{Widths of KL$_{2,3}$ atomic level for Ca, Fe, Zn}
\author{Karol Kozio{\l}$^{1,2}$} 
\email{mail@karol-koziol.net}
\affiliation{$^1$Faculty of Chemistry, Nicolaus Copernicus University, 87-100 Toru\'n, Poland}
\affiliation{$^2$Natural and Exact Science Faculty, Northeastern University of Argentina, W3404AAS Corrientes, Argentina}

\begin{abstract}
Widths of $KL_{2,3}$ atomic levels for Ca, Fe, Zn has been calculated in a fully-relativistic way using the extensive multiconfiguration Dirac-Fock and modified Dirac-Hartree-Slater calculations. The study of de-excitation of the $K^{-1}L_{2,3}^{-1}$ hole state has been presented. Additionally, the approximation to $KL_{2,3}$ level widths has been examined. 
\end{abstract}


\maketitle

\section{Introduction}

The multi-hole atomic levels are commonly present in x-ray emission spectroscopy studies. The $KL_{2,3}$ atomic level is an initial level for the transition of $K\alpha_{1,2}L_{2,3}$ satellite transitions \cite{polasik1989a} and a final level for the $K^h\alpha_{1,2}$ transitions \cite{briand}. 
In 2010, Hoszowska et al. \cite{hoszowska-new} suggested that the $\Gamma_{L_2(K)}$ level widths (i.e. the width of $\Gamma_{L_2}$ level in the presence of $K$-shell hole) is remarkably higher than the $\Gamma_{L_2}$ level width, which would explain large $K^h\alpha_{1,2}$ linewidths observed. A year later, Polasik et al. \cite{lifetimes-prl} pointed out that this statement is not essential for explaining the $K^h\alpha_{1,2}$ linewidths. In his paper, the width of the $KL_{2,3}$ level have been estimated as a sum of the widths of both $K$ and $L_{2,3}$ levels, but this estimation has not been discussed in greater details. 
The issue of the widths and the fluorescence yields linked to multi-ionized hole states has been discussed only briefly in the literature. In 1971, Larkins \cite{larkins} presented a study about the changes of the $K$-shell fluorescence yield of argon caused by additional ionization in the $L$ shell. Next, in 1995, the study of the $K$-shell fluorescence yield and level width in the presence of the $L$-shell hole for lanthanum had been carried out by Anagnostopoulos \cite{anag1995}. 

In the present work, the widths of the $KL_{2,3}$ atomic levels for Ca, Fe, Zn have been calculated in a fully-relativistic way. Additionally, the study of de-excitation of the $K^{-1}L_{2,3}^{-1}$ hole state has been presented, and the approximation to the $KL_{2,3}$ level widths considered as a sum of the $K$ and $L_{2,3}$ levels has been examined. 

\section{Theoretical background}

\subsection{The width of the $KL_{2,3}$ atomic level}

It is well established that the width of $K^{-k} L_1^{-l_1} L_2^{-l_2} L_3^{-l_3} \ldots$\ hole levels, i.e. the width of the level associated with the state having $k$ holes in the $K$ shell, $l_1$ holes in the $L_1$ subshell, etc., can be approximated by means of the following formula:
\begin{equation}
\begin{matrix}
\Gamma(K^{-k} L_1^{-l_1} L_2^{-l_2} L_3^{-l_3} \ldots) \\[1ex]
\simeq k\cdot\Gamma_K + l_1 \cdot\Gamma_{L_1} + l_2 \cdot\Gamma_{L_2} + l_3 \cdot\Gamma_{L_3} + \ldots
\end{matrix}
\end{equation}
where $\Gamma_K$ is a width of the level having one $K$-shell hole, $\Gamma_{L_1}$ is a width of the level having one $L_1$-shell hole, etc.

Hence, the width of $KL_{2,3}$ level can be approximated this way:
\begin{equation}
\Gamma_{KL} = \Gamma_{K} + \Gamma_{L}
\label{eq:klapprox}
\end{equation}

The $K^{-1}L_{2,3}^{-1}$ ($1s^{-1}2p^{-1}$) state can de-excite in the following ways (for atoms with 20$\le$Z$\le$30):\\
(a) by filling the $K$-shell hole by radiative 
$K\alpha_{1,2}L_{2,3}^{-1}$ ($KL_{2,3}$-$L_{2,3}L_{2,3}$) and $K\beta_{1,3}L_{2,3}^{-1}$ ($KL_{2,3}$-$L_{2,3}M_{2,3}$) transitions 
or non-radiative $KL_{2,3}$-$L_{2,3}LL$, $KL_{2,3}$-$L_{2,3}LM$, $KL_{2,3}$-$L_{2,3}LN$, $KL_{2,3}$-$L_{2,3}MM$, and $KL_{2,3}$-$L_{2,3}MN$ transitions;\\
(b) by filling the $L_{2,3}$-shell hole by radiative 
$L\alpha_{1,2}K^{-1}$ ($KL_3$-$KM_{4,5}$), $L\beta_{1}K^{-1}$ ($KL_2$-$KM_4$), $L\eta K^{-1}$ ($KL_2$-$KM_1$), $LlK^{-1}$ ($KL_3$-$KM_1$), $L\gamma_5K^{-1}$ ($KL_2$-$KN_1$), and $L\beta_6K^{-1}$ ($KL_3$-$KN_1$) transitions 
or non-radiative $KL_{2,3}$-$KMM$, $KL_{2,3}$-$KMN$, $KL_{2,3}$-$KNN$ (Auger type) and $KL_{2}$-$KL_{3}M$ and $KL_{2}$-$KL_{3}N$ (Coster-Kronig type) transitions;\\
(c) by Two Electron One Photon transitions (TEOP, \cite{teop1}) of less intensity.

Neglecting TEOP transitions, the total rate of transitions associated with the $K^{-1}L_{2,3}^{-1}$ state de-excitation can be expressed as:
\begin{equation}
\begin{matrix}
W_{KL} = W_{K(L)} + W_{L(K)} = \\[1ex]
W_{K(L)}^{Rad} + W_{K(L)}^{Nrad} + W_{L(K)}^{Rad} + W_{L(K)}^{Nrad} \;,
\end{matrix}
\end{equation}
where $W_{K(L)}$ is a sum of transition rates for transitions leading to filling the $K$-shell hole in the presence of the $L_{2,3}$-shell hole and $W_{L(K)}$ is a sum of transition rates for transitions which lead to filling the $L_{2,3}$-shell hole in the presence of the $K$-shell hole (next, in radiative or non-radiative way, respectively). 
The relation mentioned above merged with the relation, in which 
\begin{equation}
\Gamma = \frac{\hbar}{\tau} = \hbar W = \hbar \sum_i W_i\;, 
\label{eq:gam}
\end{equation}
where $\Gamma$ is a width of an atomic level and $W_i$ is a rate of the $i$th transition leading to de-excitation of the $K^{-1}L_{2,3}^{-1}$ state, 
leads to the relation expressed in the equation below 
\begin{equation}
\begin{matrix}
\Gamma_{KL} = \Gamma_{K(L)} + \Gamma_{L(K)} = \\[1ex]
\Gamma_{K(L)}^{Rad} + \Gamma_{K(L)}^{Nrad} + \Gamma_{L(K)}^{Rad} + \Gamma_{L(K)}^{Nrad} \;.
\end{matrix}
\label{eq:gammakl}
\end{equation}
Moreover, the fluorescence yield for the $KL_{2,3}$ level can be expressed as
\begin{equation}
\omega_{KL} = \frac{\Gamma_{K(L)}^{Rad} + \Gamma_{L(K)}^{Rad}}{\Gamma_{KL}} \;.
\label{eq:omegakl}
\end{equation}
The $K(L)$ and $L(K)$ indices in Eqs. \eqref{eq:gammakl} and \eqref{eq:omegakl} are used to mark auxiliary terms associated to the $K$ shell in the presence of the $L_{2,3}$-shell hole and the $L_{2,3}$ shell in the presence of the $K$-shell hole, respectively. 

\begin{table*}[htb!]
\centering
\caption{The widths of the $KL_{2,3}$ level compared to a sum of the $K$ and $L_{2,3}$ level widths (all values are in eV).\label{klapprox1}}
\begin{tabular*}{\linewidth}{@{\extracolsep{\fill}}l cccc cccc cc}
\hline
\multicolumn{11}{c}{~}\\[-1.5ex]
 & $\Gamma_{K(L)}^{Rad}$ & $\Gamma_{K(L)}^{Nrad}$ & $\Gamma_{K(L)}$ & $\Gamma_{K}$ & $\Gamma_{L(K)}^{Rad}$ & $\Gamma_{L(K)}^{Nrad}$ & $\Gamma_{L(K)}$ & $\Gamma_{L}$ & $\Gamma_{KL}$ & $\Gamma_{K+L}$\\[0.7ex]
\hline\\[-1.5ex]
Ca & 0.110 & 0.534 & 0.643 & 0.751 & 7.88$\times10^{-5}$ & 0.237 & 0.237 & 0.213 & 0.881 & 0.965 \\[0.5ex]
Fe & 0.361 & 0.622 & 0.983 & 1.157 & 3.59$\times10^{-3}$ & 0.453 & 0.457 & 0.405 & 1.440 & 1.562 \\[0.5ex]
Zn & 0.652 & 0.668 & 1.320 & 1.552 & 1.08$\times10^{-2}$ & 0.701 & 0.712 & 0.661 & 2.032 & 2.213 \\[1ex]
\hline
\end{tabular*}
\end{table*}

\begin{table*}[htb!]
\centering
\caption{$\Gamma_{K(L)}/\Gamma_{K}$ and $\Gamma_{L(K)}/\Gamma_{L}$ level width ratios (radiative, non-radiative, and in total) and $\Gamma_{KL}/\Gamma_{K+L}$ width ratios.\label{klapprox2}}
\begin{tabular*}{\linewidth}{@{\extracolsep{\fill}}l ccccccc}
\hline
\multicolumn{8}{c}{~}\\[-1.5ex]
 & $\displaystyle\frac{\Gamma_{K(L)}^{Rad}}{\Gamma_{K}^{Rad}}$ & $\displaystyle\frac{\Gamma_{K(L)}^{Nrad}}{\Gamma_{K}^{Nrad}}$ & $\displaystyle\frac{\Gamma_{K(L)}}{\Gamma_{K}}$ & 
$\displaystyle\frac{\Gamma_{L(K)}^{Rad}}{\Gamma_{L}^{Rad}}$ & $\displaystyle\frac{\Gamma_{L(K)}^{Nrad}}{\Gamma_{L}^{Nrad}}$ & $\displaystyle\frac{\Gamma_{L(K)}}{\Gamma_{L}}$ & 
$\displaystyle\frac{\Gamma_{KL}}{\Gamma_{K+L}}$\\[2.5ex]
\hline\\[-1.5ex]
Ca & 0.899 & 0.848 & 0.856 & 1.169 & 1.113 & 1.113 & 0.913 \\[0.5ex]
Fe & 0.903 & 0.822 & 0.850 & 1.337 & 1.127 & 1.128 & 0.922 \\[0.5ex]
Zn & 0.893 & 0.813 & 0.851 & 1.264 & 1.074 & 1.077 & 0.918 \\[1ex]
\hline
\end{tabular*}
\end{table*}

\begin{table*}[htb!]
\centering
\caption{$\omega_K$, $\omega_{K(L)}$, $\omega_L$, $\omega_{L(K)}$, and $\omega_{KL}$ fluorescence yields and $\omega_{K(L)}/\omega_K$, $\omega_{L(K)}/\omega_L$, and $\omega_{KL}/\omega_K$ fluorescence yields ratios.\label{klapprox3}}
\begin{tabular*}{\linewidth}{@{\extracolsep{\fill}}l cccccccc}
\hline
\multicolumn{9}{c}{~}\\[-2.5ex]
 & $\omega_K$ & $\omega_{K(L)}$ & $\omega_{K(L)}/\omega_K$ & $\omega_L$ & $\omega_{L(K)}$ & $\omega_{L(K)}/\omega_L$ & $\omega_{KL}$ & $\omega_{KL}/\omega_K$\\[0.7ex]
\hline\\[-1.5ex]
Ca  & 0.162 & 0.171 & 1.051 & 3.16$\times10^{-4}$ & 3.32$\times10^{-4}$ & 1.051 & 0.125 & 0.768\\[0.5ex]
Fe  & 0.346 & 0.367 & 1.063 & 6.63$\times10^{-3}$ & 7.86$\times10^{-3}$ & 1.185 & 0.253 & 0.733\\[0.5ex]
Zn  & 0.471 & 0.494 & 1.050 & 1.30$\times10^{-2}$ & 1.52$\times10^{-2}$ & 1.174 & 0.326 & 0.693\\[1ex]
\hline
\end{tabular*}
\end{table*}

\subsection{Calculation details}

The calculations of radiative transition rates have been carried out by means of \textsc{Grasp2k} code~\cite{grasp2k}. 
This code is based on the Multi-Configuration Dirac-Fock (MCDF) method. 
The presented radiative widths have been calculated by using Babushkin~\cite{cech-bab} gauge. 
In order to investigate systematic difference between radiative widths calculated by using Babushkin and Coulomb gauges, in the context of the $K$-shell level width calculation, see Ref.~\cite{kwid}. The calculations of non-radiative transition rates have been carried out by means of the \textsc{Fac} code~\cite{fac1,fac2}, which is based on a modified Dirac-Hartree-Slater (DHS) method~\cite{fac2}. 

The methodology of MCDF calculations performed in the present studies is similar to that published earlier, in many papers (see, e.g., \cite{gr1,gr2,gr6,grant1,grant2}).
The effective Hamiltonian for an $N$-electron system is expressed by
\begin{equation}
H = \sum_{i=1}^{N} h_{D}(i) + \sum_{j>i=1}^{N} C_{ij},
\end{equation}
where $h_{D}(i)$ is the Dirac operator for the $i$th electron and the terms $C_{ij}$ account for electron-electron interactions. 
The latter is a sum of the Coulomb interaction operator and the transverse Breit operator. An atomic state function (ASF) with the total angular momentum $J$ and parity $p$ is assumed in the form
\begin{equation}
\Psi_{s} (J^{p} ) = \sum_{m} c_{m} (s) \Phi ( \gamma_{m} J P ),
\end{equation}
where $\Phi ( \gamma_{m} J^{p} )$ are the configuration state functions (CSF), $c_{m}(s)$ are the configuration mixing coefficients for state $s$, and $\gamma_{m}$ represents all information required to define a certain CSF uniquely. 
In present calculations, the initial and final states of considered transitions have been optimized separately and the biorthonormal transformation has been used for performing transition rates calculations \cite{grasp2k}. Following this, the so-called relaxation effect is taken into account. 
Apart from the transverse Breit interaction, two types of quantum electrodynamics (QED) corrections (self-energy and vacuum polarization) have been included. 
On the whole, the multiconfiguration DHS method is similar to the MCDF method, but a simplified expression for the electronic exchange integrals is used \cite{fac2}. 

At this point, some of the decisions made while performing calculations recquire explanation.
Firstly, because of mixing the CSFs involving $L_2$ and $L_3$ hole states within ASFs, the width of the $KL_{2,3}$ level is taken into consideration without differentiating on the $KL_{2}$ and $KL_{3}$ levels. 
Secondly, as a consequence of coupling the initial states for the $L\alpha_{1,2}$ and $L\beta_{1}$ transitions, as well as the final states for $L\eta$ -- $Ll$ and $L\gamma_5$ -- $L\beta_6$ transitions, the transition rates for these three pairs have been calculated together. 
Thirdly, since there is a lot of atomic levels originating from a given spectator hole for open-shell atomic systems, the average width of any atomic level can be expressed by the formula 
\begin{equation}
\begin{matrix}
\Gamma = \Gamma^{Rad} + \Gamma^{Nrad} = 
\hbar \left( \sum_i \bar X_i + \sum_j \bar A_j \right) =\\
\hbar \left( \frac{1}{n}\sum\limits_{i,p,q} X_{i,p,q} + \frac{1}{n}\sum\limits_{j,r,s} A_{j,r,s} \right)\;,
\end{matrix}
\label{eq:ave}
\end{equation}
where the average transition rates, $\bar X$ and $\bar A$, are introduced, $n$ is a number of hole levels corresponding to a given hole electronic configuration, $X_{i,p,q}$ stands for transition rate for the transition between the $p$th initial level and the $q$th final level according to the $i$th de-excitation channel, and for the $A_{j,r,s}$ rate correspondingly.

\section{Results and discussion}

The widths of the $KL_{2,3}$ level ($\Gamma_{KL}$) for Ca, Fe, and Zn, compared to a sum of $K$ and $L_{2,3}$ level widths ($\Gamma_{K+L} = \Gamma_{K} + \Gamma_{L}$), are presented in Table~\ref{klapprox1}. 
Table~\ref{klapprox2} presents the collection of the values of $\Gamma_{KL}/\Gamma_{K+L}$ width ratios, as well as, the values of $\Gamma_{K(L)}/\Gamma_{K}$ and $\Gamma_{L(K)}/\Gamma_{L}$ level width ratios (radiative, non-radiative, and total). 
Table~\ref{klapprox3} deals with the values of $\omega_K$, $\omega_{K(L)}=\Gamma^{Rad}_{K(L)}/\Gamma_{K(L)}$, $\omega_L$, $\omega_{L(K)}=\Gamma^{Rad}_{L(K)}/\Gamma_{L(K)}$, and $\omega_{KL}$ fluorescence yields and $\omega_{K(L)}/\omega_K$, $\omega_{L(K)}/\omega_L$, and $\omega_{KL}/\omega_K$ fluorescence yield ratios. 

From Tables~\ref{klapprox1} and~\ref{klapprox2} one can see that: 
(a) $\Gamma_{K(L)}$ values are smaller by about 10\% than $\Gamma_{K}$ values;
(b) $\Gamma_{L(K)}$ values are larger by about 10\% than $\Gamma_{L}$ values;
(c) $\Gamma_{KL}$ values are smaller by about 8\% than $\Gamma_{K+L}$ values.
There are a few physical effects involved in the explanation of these findings. 

Statistically, removing one electron from the filled $2p$ subshell decreases by $\tfrac{1}{6}$ the probability of filling the hole in the $1s$ shell by electron jump from the $2p$ subshell (i.e. by $K\alpha_{1,2}$ transitions), because the number of $2p$-electrons being able to fill the $1s$-shell hole is reduced from 6 to 5. 
However, removing an electron from the $2p$ subshell introduces changes also in the orbitals (contraction mostly) due to changing the electron shielding. This, in turn, leads to increasing the $K\alpha_{1,2}$ and $K\beta_{1,3}$ transition rates. 
The transition rate for electric dipole (E1) transition between the initial state $\ket{i} = \ket{\gamma_{i} J_i P_i}$ and the final state $\ket{f} = \ket{\gamma_{f} J_f P_f}$ depends on transition energy $E_{tr}$ and on squared transition operator matrix element $\left|\sandwich{i}{e \vec r}{f}\right|^2$ (so-called line strength, $S_{if}$). This dependency can be expressed as (see e.g.  \cite{fval}):
\begin{equation}
W_{i \to f} = \frac{4}{3}\frac{E_{tr}^3}{\hbar c^3} \; \left|\sandwich{i}{e \vec r}{f}\right|^2 \frac{1}{d_i}
\end{equation}
where $d_i$ is the degeneracy of the initial state. 
The change of both transition energy and line strength depends on the change of the orbitals. 
For better understanding of the above expression the case of $\ket{1s^{-1}}_{J=1/2}\to\ket{3p^{-1}}_{J=3/2}$ ($K\beta_{1}$) and $\ket{1s^{-1}2p^{-1}_{3/2}}_{J=2}\to\ket{2p^{-1}_{3/2}3p^{-1}_{3/2}}_{J=3}$ (later referred to as $K\beta_{1}L_3$) transitions for Zn will be now analyzed. 
Since the $\ket{1s^{-1}2p^{-1}_{3/2}}_{J=2}$ state can be originated only from coupling of the $\ket{1s^{-1}}_{J=1/2}$ and $\ket{2p^{-1}}_{J=3/2}$ hole states and the $\ket{2p^{-1}_{3/2}3p^{-1}_{3/2}}_{J=3}$ state can be originated only from coupling of the $\ket{3p^{-1}}_{J=3/2}$ and $\ket{2p^{-1}}_{J=3/2}$ hole states, 
the $\ket{1s^{-1}2p^{-1}_{3/2}}_{J=2}\to\ket{2p^{-1}_{3/2}3p^{-1}_{3/2}}_{J=3}$ transition can be considered as the $K\beta_{1}$ transition in the presence of the $\ket{2p^{-1}}_{J=3/2}$ hole state without additional contribution of any other CSFs. 
For the clearer analysis, the reduced transition rate $\bar W_{i \to f}=d_i W_{i \to f}$ is used in order to omit the analysis of the states degeneracy. 
As one can see from Table~\ref{tab:kbl}, the transition energy, reduced rate, and line strength for the $K\beta_{1}L_3$ transition are slightly larger than it is in the case of the $K\beta_{1}$ transition. This difference can be attributed to the presence the $L_3$-shell hole. 
In other words, the $1s^{-1}2p^{-1}$ initial hole states are more excited than the $1s^{-1}$ initial hole state, the $2p^{-2}$ and $2p^{-1}3p^{-1}$ final hole states are more excited than the $2p^{-1}$ and $3p^{-1}$ final hole states, which results in $1s^{-1}2p^{-1}\to2p^{-2}$ and $1s^{-1}2p^{-1}\to2p^{-1}3p^{-1}$ transitions being of larger intensity (and energy) than $1s^{-1}\to2p^{-1}$ and $1s^{-1}\to3p^{-1}$ transitions. 

\begin{table}[!htb]
\caption{\label{tab:kbl}The case study: differences between parameters of $K\beta_{1}$ and $K\beta_{1}L_3$ transitions (see text for details) for Zn.}
\medskip
\begin{tabular*}{\columnwidth}{@{\extracolsep{\fill}}lccc}
\hline\\[-2.5ex]
 & $K\beta_{1}$ & $K\beta_{1}L_3$ & $K\beta_{1}L_3$/$K\beta_{1}$\\
 & & & ratio \\[0.5ex]\hline\\[-2.0ex]
$E_{tr}$ (eV) & 9568.849 & 9645.421 & 1.00800\\
$S$ & 1.81902$\times10^{-4}$ & 1.91432$\times10^{-4}$ & 1.05239\\
$\bar W$ (s$^{-1}$) & 1.69439$\times10^{14}$ & 1.82619$\times10^{14}$ & 1.07779\\[0.5ex]
\hline
\end{tabular*}
\end{table}


Another issue concerns the point that owing to the additional hole in the $2p$ subshell, the number of the initial and final states for $K\alpha_{1,2}$ and $K\beta_{1,3}$ transitions is higher. This influences the average $K\alpha_{1,2}$ and $K\beta_{1,3}$ transition rates (these rates do not always increase, because of the exclusion of some states due to selection rules -- see Table~\ref{tab:k-trx}).
Similar effects are observable in the study of non-radiative de-excitation processes (see Table~\ref{tab:k-tra}) 
in the context of filling the $K$ shell. 

All of the above findings show that removing one electron from the $2p$ subshell decreases the radiative width of the $K$ shell  by less than $\tfrac{1}{6}$. 
The results for $\Gamma_{K(L)}$-to-$\Gamma_{K}$ ratio gathered in the present work are close to those obtained by Anagnostopoulos for lanthanum~\cite{anag1995}. 

For de-excitation processes linked to filling the $2p$-shell hole, there are only the effects of increasing the radiative or non-radiative transition rates, which result from changes of orbitals and numbers of states (see Tables~\ref{tab:l-trx} and ~\ref{tab:l-tra}). 
As the result, the width of the $L_{2,3}$ shell increases in the presence of the additional $K$-shell hole. 

These two effects of opposite outcome, i.e. decreasing the width of the $K$ level in the presence of the $L_{2,3}$-shell hole and increasing the width of the $L_{2,3}$ level in the presence of the $K$-shell hole, do not cancel each other completely. 
As the result, the width of the $KL_{2,3}$ atomic level is slightly smaller than a sum of widths of the $K$ and $L_{2,3}$ atomic levels. 

Table~\ref{klapprox3} illustrates that the differences between the fluorescence yields for $K$-shell hole level and the ones for the $K$-shell hole level in the presence of the $L_{2,3}$ hole are about 5-6\%. 
These values varies from the difference values obtained by Anagnostopoulos for lanthanum~\cite{anag1995}, which do not exceed 1\%, and from the value obtained by Larkins for argon~\cite{larkins}, which is equal to 10\%. 
These combined results show clearly $Z$-dependence. 
Moreover, it is worth noting that the fluorescence yield for the $KL_{2,3}$ level are 23-31\% smaller than the fluorescence yield for the $K$ level for considered elements. 

\section{Conclusions}
The detailed relativistic calculations for the width of the $KL_{2,3}$ level for Ca, Fe, and Zn have been presented for the first time. 
Basing on the findings presented above, two main conclusions can be drawn: 
(a) the assumption stated in the work of Polasik et al. \cite{lifetimes-prl}, concerning the approximation of the width of the $KL_{2,3}$ level as a sum of widths of $K$ and $L_{2,3}$ levels, proves to be justified for atoms considered in therein work, as the approximation errors (i.e. the differences between the width of the $KL_{2,3}$ level and a sum of the widths of $K$ and $L_{2,3}$ levels) are smaller than the other effects considered in the work of Polasik et al. \cite{lifetimes-prl}; 
(b) the suggestion that the width of the $L_2$ level in the presence of $K$-shell hole is much larger than the $L_2$ level width without the presence of $K$-shell hole, stated in the work of Hoszowska et al. \cite{hoszowska-new}, is not valid. 
The results presented above may be of a great help in interpreting and benchmarking the $K$-shell satellite, $K\alpha_{1,2}L_{2,3}$, and the hypersatellite, $K^h\alpha_{1,2}$, x-ray spectra measured with high resolution. 
Moreover, the Author of present work hopes that these findings will be also helpful in general discussion about the widths of multi-hole atomic levels, in the view of the fact that multi-hole states are commonly present in x-ray emission spectroscopy studies.

\subsection*{Acknowledgments}
The author is thankful to Gustavo Aucar (Northeastern University of Argentina) for his hospitality while this paper was being prepared.

\begin{table*}[!htb]
\caption{\label{tab:k-trx}Averaging $K\alpha_{1,2}$ and $K\beta_{1,3}$ (de-excitation of $K^{-1}$ states) and $K\alpha_{1,2}L_{2,3}^{-1}$ and $K\beta_{1,3}L_{2,3}^{-1}$ (de-excitation of $K^{-1}L_{2,3}^{-1}$ states) transition rates for Ca, Fe, Zn.}
\medskip
\begin{tabular*}{\linewidth}{@{\extracolsep{\fill}}c cccc}
\hline\\[-2.5ex]
 & \multicolumn{4}{c}{$\bar X_i$ (s$^{-1}$)}\\[0.5ex]
\cline{2-5}\\[-2ex]
 & $K\alpha_{1,2}$ ($\times10^{14}$) & $K\beta_{1,3}$ ($\times10^{13}$) & $K\alpha_{1,2}L_{2,3}^{-1}$ ($\times10^{14}$) & $K\beta_{1,3}L_{2,3}^{-1}$ ($\times10^{13}$)\\[0.5ex]\hline\\[-2.0ex]
Ca & 1.579 & 2.758 & 1.570 & 0.984 \\
Fe & 5.234 & 8.411 & 4.528 & 9.582 \\
Zn & 9.814 & 12.859 & 9.118 & 7.926 \\[0.5ex]
\hline
\end{tabular*}
\end{table*}

\begin{table*}[!htb]
\caption{\label{tab:k-tra}Averaging $K$-$LL$, $K$-$LM$, $K$-$LN$, $K$-$MM$, and $K$-$MN$ (de-excitation of $K^{-1}$ states) and $KL_{2,3}$-$L_{2,3}LL$, $KL_{2,3}$-$L_{2,3}LM$, $KL_{2,3}$-$L_{2,3}LN$, $KL_{2,3}$-$L_{2,3}MM$, and $KL_{2,3}$-$L_{2,3}MN$ (de-excitation of $K^{-1}L_{2,3}^{-1}$ states) transition rates for Ca, Fe, Zn.}
\medskip
\begin{tabular*}{\linewidth}{@{\extracolsep{\fill}}c ccccc ccccc}
\hline\\[-2.5ex]
 & \multicolumn{10}{c}{$\bar A_i$ (s$^{-1}$)}\\[0.5ex]
\cline{2-11}\\[-2ex]
 & $K$-$LL$ & $K$-$LM$ & $K$-$LN$ & $K$-$MM$ & $K$-$MN$ & $KL_{2,3}$-$L_{2,3}LL$ & $KL_{2,3}$-$L_{2,3}LM$ & $KL_{2,3}$-$L_{2,3}LN$ & $KL_{2,3}$-$L_{2,3}MM$ & $KL_{2,3}$-$L_{2,3}MN$ \\[0.5ex]
 & ($\times10^{14}$) & ($\times10^{14}$) & ($\times10^{12}$) & ($\times10^{13}$) & ($\times10^{11}$) & ($\times10^{14}$) & ($\times10^{14}$) & ($\times10^{12}$) & ($\times10^{13}$) & ($\times10^{11}$) \\[0.5ex]\hline\\[-2.5ex]
Ca & 7.744 & 1.677 & 4.388 & 0.918 & 5.318 & 6.156 & 1.761 & 5.804 & 1.229 & 8.658 \\
Fe & 9.165 & 2.166 & 4.072 & 1.274 & 5.296 & 7.145 & 2.103 & 5.046 & 1.484 & 7.604 \\
Zn & 9.861 & 2.428 & 3.821 & 1.468 & 5.096 & 7.618 & 2.309 & 4.648 & 1.656 & 7.089 \\[0.5ex]
\hline
\end{tabular*}
\end{table*}

\begin{table*}[!htb]
\caption{\label{tab:l-trx}Averaging $L\alpha_{1,2}$+$L\beta_{1}$, $L\eta$+$Ll$, and $L\gamma_5$+$L\beta_6$ (de-excitation of $L_{2,3}^{-1}$ states) and $L\alpha_{1,2}K^{-1}$+$L\beta_{1}K^{-1}$, $L\eta K^{-1}$+$LlK^{-1}$, and $L\gamma_5K^{-1}$+$L\beta_6K^{-1}$ (de-excitation of $K^{-1}L_{2,3}^{-1}$ states) transition rates for Ca, Fe, Zn.}
\medskip
\begin{tabular*}{\linewidth}{@{\extracolsep{\fill}}c ccc ccc}
\hline\\[-2.5ex]
 & \multicolumn{6}{c}{$\bar X_i$ (s$^{-1}$)}\\[0.5ex]
\cline{2-7}\\[-2ex]
 & $L\alpha_{1,2}$+$L\beta_{1}$ & $L\eta$+$Ll$ & $L\gamma_5$+$L\beta_6$ & $L\alpha_{1,2}K^{-1}$+$L\beta_{1}K^{-1}$ & $L\eta K^{-1}$+$LlK^{-1}$ & $L\gamma_5K^{-1}$+$L\beta_6K^{-1}$ \\[0.5ex]
 & ($\times10^{12}$) & ($\times10^{11}$) & ($\times10^{10}$) & ($\times10^{12}$) & ($\times10^{11}$) & ($\times10^{10}$) \\[0.5ex]\hline\\[-2.0ex]
Ca & 0.000 & 0.953 & 0.701 & 0.000 & 1.084 & 1.133 \\
Fe & 3.714 & 3.492 & 1.686 & 5.055 & 3.724 & 2.542 \\
Zn & 12.327 & 6.780 & 2.795 & 15.735 & 7.038 & 4.073 \\[0.5ex]
\hline
\end{tabular*}
\end{table*}

\begin{table*}[!htb]
\caption{\label{tab:l-tra}Averaging $L_{2,3}$-$MM$, $L_{2,3}$-$MN$, $L_{2,3}$-$NN$, $L_{2}$-$L_{3}M$, and $L_{2}$-$L_{3}N$ (de-excitation of $L_{2,3}^{-1}$ states) and $KL_{2,3}$-$KMM$, $KL_{2,3}$-$KMN$, $KL_{2,3}$-$KNN$, $KL_{2}$-$KL_{3}M$, and $KL_{2}$-$KL_{3}N$ (de-excitation of $K^{-1}L_{2,3}^{-1}$ states) transition rates for Ca, Fe, Zn.}
\medskip
\begin{tabular*}{\linewidth}{@{\extracolsep{\fill}}c ccccc}
\hline\\[-2.5ex]
 & \multicolumn{5}{c}{$\bar A_i$ (s$^{-1}$)}\\[0.5ex]
\cline{2-6}\\[-2ex]
 & $L_{2,3}$-$MM$ ($\times10^{14}$) & $L_{2,3}$-$MN$ ($\times10^{12}$) & $L_{2,3}$-$NN$ ($\times10^{10}$) & $L_{2}$-$L_{3}M$ ($\times10^{8}$) & $L_{2}$-$L_{3}N$ ($\times10^{13}$) \\[0.5ex]\hline\\[-2.0ex]
Ca & 3.151 & 8.834 & 3.659 & 0.000 & 0.000 \\
Fe & 6.027 & 7.094 & 2.106 & 6.078 & 0.100 \\
Zn & 9.620 & 6.333 & 1.528 & 0.000 & 2.266 \\[0.5ex]
\hline\\[-2.5ex]
 & \multicolumn{5}{c}{$\bar A_i$ (s$^{-1}$)}\\[0.5ex]
\cline{2-6}\\[-2ex]
 & $KL_{2,3}$-$KMM$ ($\times10^{14}$) & $KL_{2,3}$-$KMN$ ($\times10^{12}$) & $KL_{2,3}$-$KNN$ ($\times10^{10}$) & $KL_{2}$-$KL_{3}M$ ($\times10^{8}$) & $KL_{2}$-$KL_{3}N$ ($\times10^{13}$) \\[0.5ex]\hline\\[-2.0ex]
Ca & 3.486 & 11.903 & 6.062 & 0.000 & 0.000 \\
Fe & 6.790 & 9.132 & 3.314 & 0.000 & 0.008 \\
Zn & 10.400 & 8.105 & 2.386 & 0.000 & 1.644 \\[0.5ex]
\hline
\end{tabular*}
\end{table*}

%

%

\end{document}